# Systematic electrochemical etching of various metal tips for tunneling spectroscopy and scanning probe microscopy

Jiawei Zhang[1], Pinyuan Wang[1], Xuao Zhang[1], Haoran Ji[1], Jiawei Luo[1], He Wang[2], Jian Wang[1,3,4,5,a]

[1] International Center for Quantum Materials, School of Physics, Peking University, Beijing 100871, China.
[2] Department of Physics, Capital Normal University, Beijing 100048, China.
[3] Collaborative Innovation Center of Quantum Matter, Beijing 100871, China.
[4] CAS Center for Excellence in Topological Quantum Computation, University of Chinese Academy of Sciences, Beijing 100190, China.
[5] Beijing Academy of Quantum Information Sciences, Beijing 100193, China.
[a] Author to whom correspondence should be addressed: jianwangphysics@pku.edu.cn

**ABSTRACT**

Hard point-contact spectroscopy and scanning probe microscopy/spectroscopy are powerful techniques for investigating materials with strong expandability. To support these studies, tips with various physical and chemical properties are required. To ensure the reproducibility of experimental results, the fabrication of tips should be standardized, and a controllable and convenient system should be set up. Here a systematic methodology to fabricate various tips is proposed, involving electrochemical etching reactions. The reaction parameters fall into four categories: solution, power supply, immersion depth, and interruption. An etching system was designed and built so that these parameters could be accurately controlled. With this system, etching parameters for copper, silver, gold, platinum/iridium alloy, tungsten, lead, niobium, iron, nickel, cobalt, and permalloy were explored and standardized. Among these tips, silver and niobium's new recipes were explored and standardized. Optical and scanning electron microscopies were performed to characterize the sharp needles. Relevant point-contact experiments were carried out with an etched silver tip to confirm the suitability of the fabricated tips.

## I. INTRODUCTION

As a powerful tool to study superconductors, the point-contact spectroscopy (PCS) technique has been successfully applied to the investigation of materials with various properties.[1-8] In experiments, PCS is categorized as soft point-contact and hard point-contact. The former normally utilizes silver paint to form the point-contact. The usage of tips in hard point-contact endows PCS with more possibilities. Traditionally, one may conveniently measure the superconducting gap $\Delta$ and pairing symmetry of a superconductor, as well as energy information about quasiparticle excitations such as magnons and phonons, via PCS.[1-5] In recent years, tip-induced or -enhanced superconductivity has been found in hard point-contact experiments, whose mechanism is attributed to a local doping effect, local high pressure effect, and interface effect on the boundary.[7-13]

Various study purposes call for various metal tips. To detect the intrinsic properties of a sample,

soft metal tips are preferable.[3, 7] To introduce local high pressure or to pierce through the segregated layer on a sample's surface, hard metals like tungsten can be put to use.[7, 9, 10] Tips made of superconductors act as probes of a sample's spin-polarization,[14] and ferromagnetic tips can be utilized to detect the pairing symmetry of superconductors.[8, 15] A range of properties of a sample can be determined by applying different tips as probes, as well as by introducing control experiments between soft and hard, ferromagnetic and paramagnetic, normal and superconducting tips. All these applications require a standard fabrication method for refined tips, with reproducibility and convenience of handling.

The tip's needle curvature is one of the essential quality parameters. In hard point-contact, we require the needle curvature radius to have a magnitude of hundreds of nanometers, to avoid the loss of energy or momentum due to inelastic scattering in the contact area, and to ensure PCS gets energy-well-resolved spectra.[5, 13] Besides, in scanning probe microscopy (SPM) techniques such as scanning tunneling microscopy/spectroscopy (STM/S), superconducting and ferromagnetic tips are also utilized.[16-18] Tungsten, iron and cobalt tips also serve various purposes in atomic force microscopy (AFM) systems.[19]

Electrochemical etching methods for several individual kinds of tips have been reported in previous literature, including gold[20] and tungsten,[21] which have been widely used in SPM and tunneling microscopy, including STM/S and PCS. Compared with electrochemical etching, mechanical methods of tip fabrication, including polishing and shearing, have some disadvantages. One is that the radii of tips are not uniform, which makes reproducibility less likely. The other is the possibility of introducing more impurities on the surface of tips in the polishing process. Thus, electrochemical etching is a preferable method due to its standardization and reproducibility.

However, the number of available tips fabricated with electrochemical etching is still limited. Here a systematic etching methodology is proposed for various tips for hard point-contact experiments, using:
- Soft metals including copper (Cu), silver (Ag) and gold (Au).
- Tungsten (W), one of the hardest metals.
- Alloy of platinum/iridium (Pt/Ir), famous for its high chemical stability.
- Superconductors including lead (Pb) and niobium (Nb).
- Common ferromagnetic metals including iron (Fe), nickel (Ni) and cobalt (Co).

The proposed methodology was suitable for all these metals, simply using a change of recipe in the developed etching system. All the tips fabricated with this methodology are indicated in Figure 1a. Among these metal tips, more convenient and less hazardous etching recipes have been developed for Ag tip and Nb tip, respectively. These tips with a fine sharp needle shape, confirmed by optical microscope and scanning electron microscope (SEM), have potential to be utilized in hard point-contact and SPM experiments.

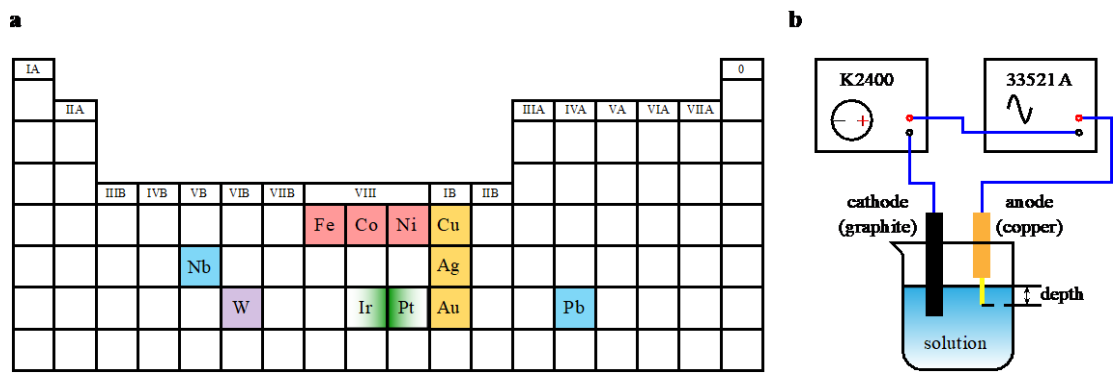

FIG. 1. Metals that are available for electrochemical etching for tips, and the etching circuit. (a) The tips that can be fabricated with the developed etching system, shown in the periodic table of the elements. Different colors represent tips' different usages in PCS experiments, such as yellow for soft tips, blue for superconducting tips and red for ferromagnetic tips. Pt and Ir are always used together in an alloy. (b) The circuit that supplies power and monitors the current. The anode holder is made of copper and the cathode is graphite.

## II. METHOD AND RESULTS

### A. Development of etching system

Electrochemical etching is based on the principle of electrochemical reaction. When a direct current (DC) is supplied to a solution, the atoms on the surface of the anodic metal lose their electrons, become cations (positive ions) and then dissolve in the solution. The overall setup is shown in Figure 1b. To avoid additional reactions on the surface of the cathode, inert graphite was chosen as the cathode. The copper marked as the anode mechanically fixed and electrically connected to the metal wire, and was not involved in the reaction.

There were four key parameters for tip fabrication: solution, power supply, immersion depth, and interruption.

For the solution, one must decide which solute to use and its concentration. For metals, especially those with reactivity higher than hydrogen, an acid environment is preferred, since hydrogen ions help with the dissolution process. Other conditions, such as the possible occurrence of reaction products with low solubility, must be considered as well. Concentration of the solute determines the reaction rate chemically, and also determines the electrical conductance of the solution. The conductance needs to be stable in order to fabricate tips with a fine shape; it was sometimes found necessary to decrease the reagents' concentration for a lower reaction rate, then introduce a salt with high solubility like potassium chloride (KCl), to maintain the stability of conductance.

The power output influences the reaction rate, and can be precisely controlled. The voltage supplied to the electrolytic cell determines whether a reaction can happen, with respect to the standard electrode potential. Thus, for given reagents and their concentration, the voltage range is fixed, and the reaction rate cannot be controlled by varying the voltage. If the voltage is too low, the reaction cannot happen. Therefore, an alternating current (AC) power supply was used instead of DC. The AC used here included both square and sine waves. By changing the duty cycle of the square wave, i.e. the proportion of high-level voltage in a period, the average supplied power could be varied without interrupting the reaction. The etching circuit is shown in Figure 1b. An arbitrary waveform generator 33521A supplied various forms of AC power, and sourcemeter K2400 supplied DC bias as well as monitored the current.

The most important parameter while preparing the reaction was found to be the immersion depth of the metal wire before the reaction began, as indicated in Figure 1b. The formation of a tip with a sharp needle shape was attributed to the effects of surface tension and gravity on the immersed part. The phenomenological schematic diagrams of reaction are shown in Figure 2. Surface tension caused the formation of convexity where the metal wire met the solution surface. This "neck" part dissolved more quickly than the immersed part (see Figure 2a, where convexity and "neck" are marked), since the current density here was the highest. With the "neck" becoming thinner as shown in Figure 2b, the immersed "bottom" eventually fell away under gravity. The thin "neck" could be observed as the intermediate by switching off the power before the "bottom" fell away (see supplementary Figure S1), proving our hypothesis about reaction process in figure 2b. The original immersion depth determined the weight of the immersed "bottom". In Figure 2f, a tip/anode holder was designed with tunable vertical position, enabling fine adjustment ($z$) of immersion depth. Rotating the handle by 1 mm (the smallest division) drove the $z$-axis motion by 0.025 mm. Thus, the precision of immersion depth equals 0.025 mm. Six screw holes on the platform allow the spacing variation between anode and cathode.

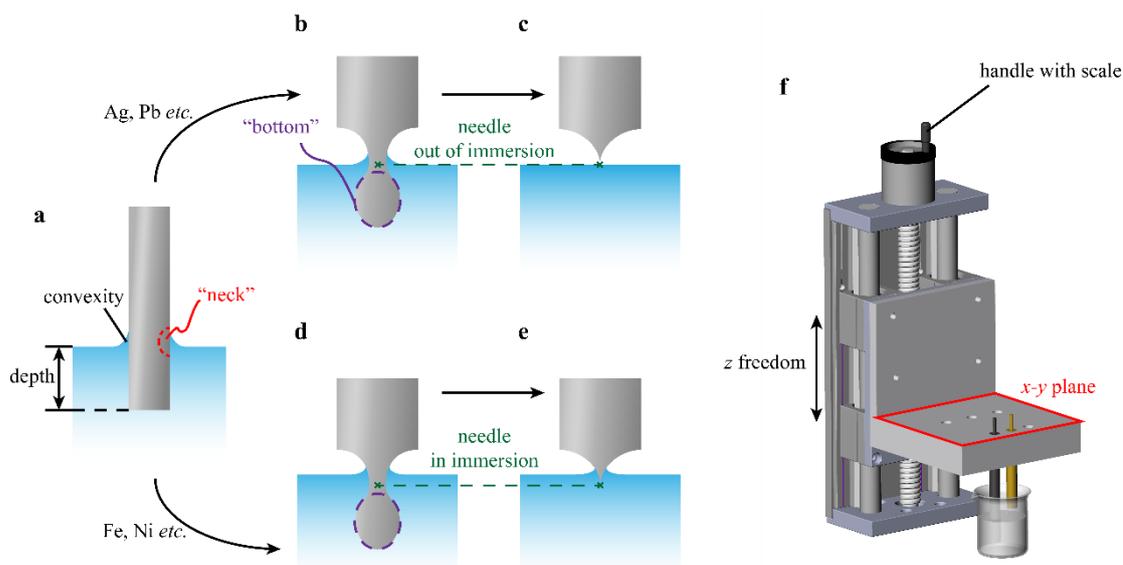

FIG. 2. Schematic diagrams of two conditions of the reaction process, and setup of the electrolytic cell. (a) A metal wire is immersed with convexity surrounding its "neck". a→b→c and a→d→e represent reactions of Ag, Pb *etc.*, and Fe, Ni *etc.*, respectively. For (b) and (d), the "bottom" parts drop under gravity, and the two conditions differ in the position of the "necks", where the fine needle shapes form. The green dashed lines show correspondence between "necks" and fine needle shapes. In (c) and (e), fine needle shapes form, one out of the solution and one immersed in it. (f) shows a specially designed etching platform. The handle with scale enables precise control of the anode's $z$ position via a worm structure. Six screw holes on the $x$-$y$ plane enable the distance between anode and cathode to be varied.

For successful fabrication, it was necessary to stop the reaction in time. Depending on the types of metal, once the convex shape formed, two possible cases occurred. For some metals such as Ag and Pb, the "neck" formed at or above the surface of the solution, so once the "bottom" fell away, the neck was above the solution and the reaction ceased, illustrated as a→b→c in Figure 2. It was not necessary to manually interrupt the reaction, which would end with the convexity vanishing. But for others such as Fe and Ni, the "neck" remained in contact with the solution even after the "bottom" fell away, so the reaction was able to continue and the needle became over-etched as shown in Figure 2a→d→e; to prevent this, the reaction could be stopped by turning off the power after the "bottom" fell off. At the moment the "bottom" fell off, the electrochemical current dropped sharply since the effective contact area of reaction had decreased; this could be used as the signal to turn off the power and end the reaction. Since K2400 in Figure 1b was used to monitor the current, we could shut off the power supply immediately after the drop of current.[20]

After the reaction was terminated, small salt crystals were sometimes found attached to the needle. In most cases they could be washed away with flowing water without breaking the fine shape of needle. Or the needle could be immersed in a beaker of water for ultrasonic cleaning for several seconds.

The detailed and standardized fabrication process of various tips are shown below. Table I shows the reaction parameters. Uncertainties of solution concentration are estimated from the uncertainties of solute mass and solvent volume during solution preparation. Square waves' low level and high level are shown in the table. The square waves mentioned here are all of duty cycle (the proportion of high-level voltage in a period) of 50%. For sine waves, $V_{rms}$ means root mean square of voltage. The mechanical precision of immersion depth equals 0.025 mm, and the uncertainties can be estimated as 0.05 mm with the operators' error considered. "Naturally" means the needle shape formed without manual intervention and "shutting after drop" means the power supply was shut off immediately after the electrochemical current dramatically dropped. Pt/Ir etching required a lengthy reaction. Needle diameters were measured with an optical microscope, except for Ag, Nb and W which are confirmed with SEM images (see Figure 4, Figure 5, and Figure S3, Figure S4 and Figure S5 of the supplemental materials). During optical and SEM measurements, the top part of the needle could be seen as a spherical cap, and the needle's curvature radius was defined by the cap's curvature radius.

TABLE I. Parameters for available tips

| Metal | Solution | Power supply | Immersion depth (mm) | How the reaction ends | Needle radius (μm) |
| --- | --- | --- | --- | --- | --- |
| Cu | 36.5% (saturated) HCl | 5.0 V DC | 1.00 | naturally | < 1.3 |
| Ag | (6.0 ±0.4) mol/L $HNO_3$ | 0–2.5 V square wave | 0.50 | naturally | ~ 0.115 (SEM) |
| Au | 20% ±1% HCl | 0–2.5 V square wave | 0.50 | naturally | < 0.3 |

| Pt/Ir | 1–2 mol/L CaCl$_2$ & a little HCl | 20 V$_{rms}$ sine wave | 1.00 | lasting for 2 h | ~ 6.7 |
|---|---|---|---|---|---|
| W | (4.0 ±0.3) mol/L NaOH | 5.0 V DC | 1.00 | shutting after drop | ~ 0.037 (SEM) |
| Pb | (2.5 ±0.2) mol/L CH$_3$COOH | 5.0 V DC | 2.50 | naturally | ~ 5.0 |
| Nb | (15 ±1) mol/L NaOH | −5.0–5.0 V square wave | 0.25 | shutting after drop | ~ 0.172 (SEM) |
| Fe | (1.3 ±0.1) mol/L KCl & (1.2 ±0.1) mol/L HCl | 0–3.0 V square wave | 0.50 | shutting after drop | < 1.3 |
| Co | (2.3 ±0.2) mol/L KCl & (2.4 ±0.2) mol/L HCl | 3.0 V DC | 0.75 | shutting after drop | < 0.8 |
| Ni | (2.10 ±0.15) mol/L KCl | 0–3.5 V square wave | 0.50 | shutting after drop | < 0.8 |
| Ni/Fe | (1.3 ±0.1) mol/L KCl & (1.2 ±0.1) mol/L HCl | 3.5 V DC | 0.50 | shutting after drop | < 1.6 |

### B. Cu, Ag, and Au tips

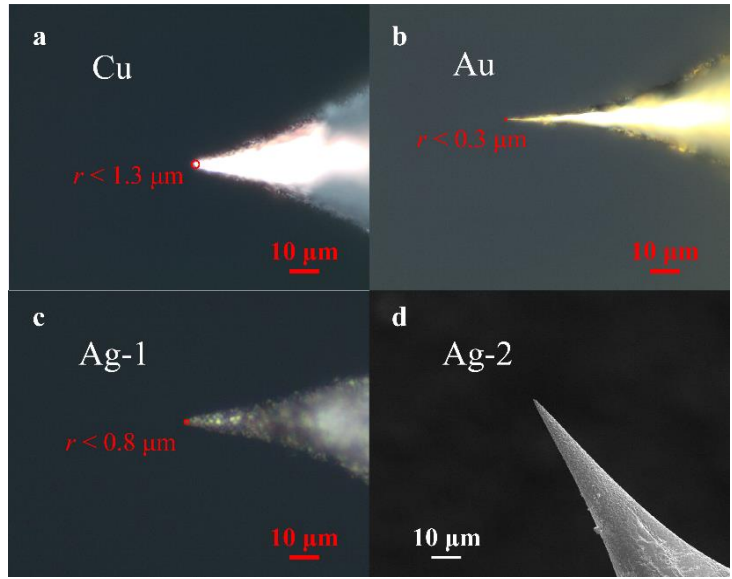

FIG. 3. Optical images of etched (a) Cu tip, (b) Au tip, and (c) Ag tip (Ag-1). (d) A SEM image of a Ag tip (Ag-2). Ag-1 and Ag-2 are two Ag tips etched with the same recipe. Scale bars are marked in the panels. The red circles are placed to best fit on the spherical cap, whose radii are regarded as curvature radii of needles.

Cu, Ag, and Au have lower reactivity than hydrogen, which prevent them from directly displacing hydrogen from an acid solution. However, these metals are easy to dissolve under current. The method for etching Au with hydrochloric acid (HCl) has been proposed previously;[20] here we give the modified parameters in Table I. In comparison with Au, Cu has similar fabrication parameters since they share a similar electrochemical reaction mechanism.

For Ag, the situation differs. Ions of Ag, i.e. $Ag^+$, can react with several anions to produce precipitation, including AgCl and $Ag_2SO_4$, which may cover the anodic metal and prevent further reaction. Thus, HCl and sulphuric acid ($H_2SO_4$) did not work for etching Ag. Since $Ag^+$ can coexist with $NO_3^-$ in the solution, nitric acid ($HNO_3$) was innovatively used to etch Ag. Etching of Ag tip in previous work involved the high voltage of AC 100 V and citric acid,[22] and our recipe is more convenient in common laboratories. Figure 3 gives the optical images of the etched Cu, Ag and Au tips, whose needle radii are all of sub-micrometer level. For Ag tip, we utilized SEM to acquire higher resolution image and accurately measure the needle curvature (Figure 3d and Figure S3c). Its curvature radius of 115 nm is better than previous results[22], which enables good performance in the PCS experiment.

## C. Superconducting tips (Pb and Nb)

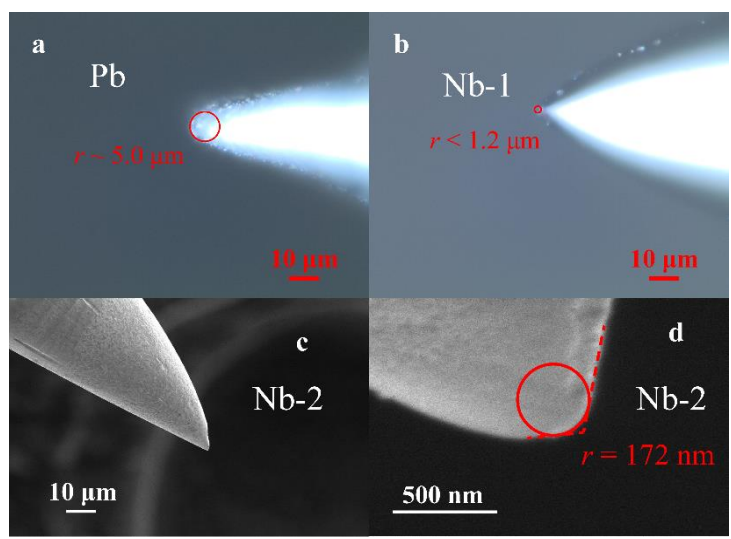

FIG. 4. Optical images of (a) Pb tip and (b) Nb tip (Nb-1). (c) A SEM image of a Nb tip (Nb-2). (d) Zoom-in of the tip in (c). Nb-1 and Nb-2 are two Nb tips etched with the same recipe. Scale bars are marked in the panels. In (d), the red circle is tangent to two red dashed lines, showing the curvature of the tip.

Superconducting tips have been used in STM/S and PCS to study superconductivity and spin polarization.[14,16,18] To further explore new superconducting tips, we developed a standard tip fabrication technique for Pb and Nb. For superconductors with $T_c$ lower than 1 K, the available measurement range of temperature would be narrow. Pb, with $T_c = 7.2$ K, and Nb, with $T_c = 9.2$ K, were therefore chosen. For Pb, difficulties were encountered similar to Ag, since many Pb salts have low solubility. We found lead acetate [$Pb(CH_3COO)_2$] to have high solubility and chose acetic acid ($CH_3COOH$) for etching. Although acetic acid is classified as a weak acid, the reaction led to a good result. The needle is shown in Figure 4a; the radius is approximately 5 μm.

Nb is difficult to etch owing to its chemical stability. Previous work used hydrofluoric acid (HF)[23] to proceed the reaction. Here the far less hazardous NaOH was used, to avoid the potential risks. In a highly concentrated solution of NaOH (15 mol/L), the reaction of Nb is described by

the following equation:
$$Nb - 5e^- + 4OH^- \rightarrow NbO_3^- + 2H_2O$$

The solution was heated to accelerate the reaction. When heated to about 85 °C, the typical reaction time reduced from about 10 minutes to about 5 minutes. The power supply was a -5.0 V– 5.0 V square wave with a duty cycle of 50%, which is an alternating waveform. SEM was also utilized to confirm the Nb tip's needle curvature radius clearly. A curvature radius of 172 nm was indicated in Figure 4d. The surface of Nb tip was smooth even in the high magnification (Figure 4d), which could be due to the relatively long etching time. In previous STM/S[16,18] and PCS[14] studies, the tips were mechanically sharpened to a curvature radius of tens of micrometers. Our process generated superconducting tips with a curvature radius of about 100 nanometers, suitable for relative STM/S and PCS studies.

**D. Pt/Ir and W tips**

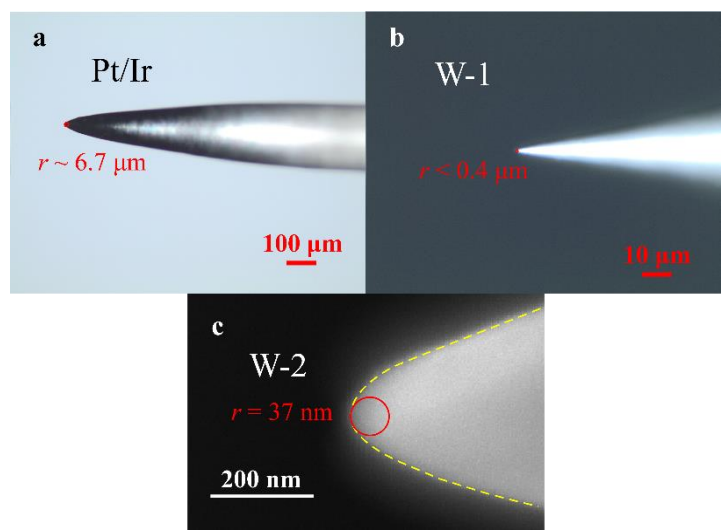

FIG. 5. Optical images of (a) Pt/Ir tip and (b) W tip (W-1). (c) A SEM image of a W tip (W-2). W-1 and W-2 are two W tips etched with the same recipe. Scale bars are marked in the panels.

Pt/Ir and W tips are relatively hard. Since Pt/Ir tips are widely used in STM, there have already been several successful trials developing techniques for Pt/Ir-tip etching.[24-29] Although softer and more expensive than tungsten, Pt/Ir is preferable for STM tips on account of its stability against oxidation, but this also means special methods are needed to accelerate the etching process. In the literature, it is widely accepted to use the $CaCl_2/HCl/H_2O$ system as the electrolyte[25-27] because formation of the complex ion $[PtCl_4]^{2-}/[PtCl_6]^{2-}$ is chemically preferred. Many other trials with solutions of strong acids such as saturated hydrochloric acid, nitric acid or their mixture—*aqua regia*—as electrolyte turned out to be failures.[28] As for the power supply, first, it is required to be alternating. Since the platinum electrode can easily be passivated, the chemical potential has to alternate to keep the surface reactive.[28] Second, if the power supply is sine wave without DC bias, its amplitude needs to be high, up to 20–25 $V_{rms}$, to accelerate the overall reaction. In our experiments, we used the $CaCl_2/HCl/H_2O$ system as the electrolyte. To find the optimal electrolyte concentration and power supply amplitude, we compared the results with various parameters. It turned out that the etching rate was fastest when the concentration of $CaCl_2$ was between 1 and 2 mol/L and the sine wave amplitude was 20 $V_{rms}$. Even with the above modifications, the etching process was still extremely time-consuming, typically one or two hours. Figure 5a shows the optical image of an etched Pt/Ir tip. Since a mechanically sharpened Pt/Ir tip can perform well in point-contact experiments,[30,31] whose curvature radius approximately equals tens of micrometers, the etched Pt/Ir tip can also perform well in point-contact experiments. This is because the effectively electric contact radius could be much smaller than the curvature radius during approaching-contacting process.[5] Compared with the previous etching results in literature for Pt/Ir tip,[27,28] our etched tip's apex angle is a little larger, ensuring its robustness in tip-induced or -enhanced superconductivity experiments.[13]

As a material that has long been used in STM/S, the fabrication of W tips is well-explored,[21,29] including thermal treatment[32] and programmable procedure[33]. Our etching parameters are the same as standard ones, with participation of sodium hydroxide (NaOH). More details can be found in Table I. According to SEM characterization (Figure 5c and Figure S5), the tips' quality is as good as results in previous reports[33], with evenly etched surfaces and needles as sharp as 37 nm. Our W tips performed well in PCS, especially experiments of tip-induced or -enhanced superconductivity.[7,10]

### E. Ferromagnetic tips (Fe, Co, Ni, and permalloy)

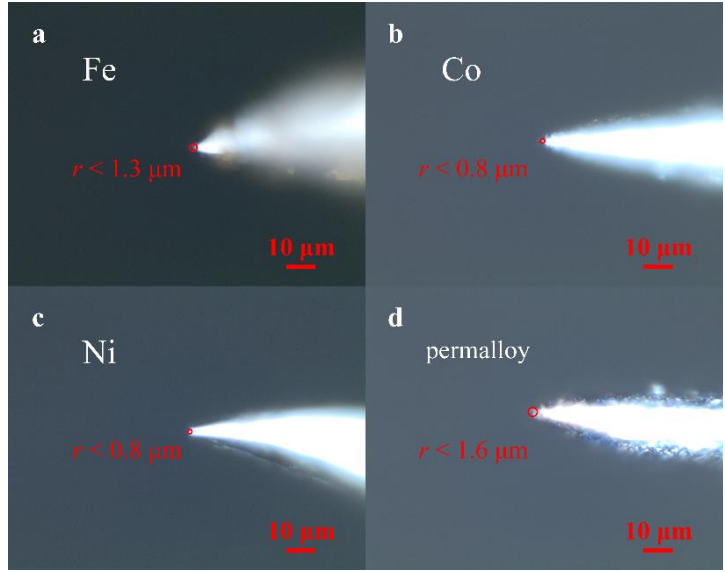

FIG. 6. Optical images of (a) Fe tip, (b) Co tip, (c) Ni tip, and (d) permalloy tip. Scale bars are marked in the panels.

Ferromagnetic tips offer a convenient platform to explore the spin-polarized transport process[15], as well as force sensing in AFM.[19] Here we present the results of Fe[34], Co[35], Ni[34], and permalloy (Ni/Fe, composed 80% of Ni and 20% of Fe). The etching of VIII group in the periodic table of the elements is rather simple, since Fe, Co, and Ni all have reactivity higher than hydrogen. But this means they can easily be over-etched, and so the rate of reaction must be reduced. In this work, the acidity was reduced by decreasing the concentration of HCl. This reduced the conductance, so additional neutral salt KCl was introduced. Note that in the etching of Ni, no HCl was needed. Another subtle factor is the end of reaction. For Fe, Co, and permalloy, the reaction needed to be terminated immediately when the current dropped, otherwise the needle became over-etched. The ferromagnetic tips fabricated with our etching process can be found in Figure 6.

## III. PERFORMANCE OF THE ETCHED TIPS

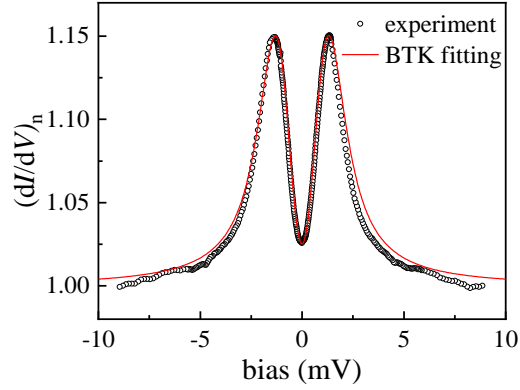

FIG. 7. Point-contact differential conductance spectra measured with an electrochemically etched Ag tip in contact with a niobium sample. Black circles stand for experimental data at 3.2 K, which is normalized by 10 K normal-state spectra. The red line represents BTK fitting, with $Z = 0.665$, $\Delta = 1.11$ meV and $\Gamma = 0.49$ meV.

To examine whether the tips fabricated in our electrochemical etching system were suitable for PCS, a low-temperature PCS experiment was performed using the etched Ag tip. Note that the innovative electrochemical etching method with $HNO_3$ involved for Ag tips has never been reported before. It was made to touch the superconductor niobium to form a point contact in a Leiden dilution refrigerator CF450, achieved by attocube's piezo stepper positioner. Figure 7 shows the point-contact spectrum with obvious double peak structure at 3.2 K, which is a typical superconducting spectrum. The spectrum is normalized by the 10 K non-superconducting spectrum background, which can be found in Figure S2b. Using Blonder-Tinkham-Klapwijk (BTK) theory,[36,37] we fitted the data with superconducting gap $\Delta = 1.11$ meV, smearing factor $\Gamma = 0.49$ meV and interface barrier $Z = 0.665$. The fitting result is consistent with the previous literature,[6] and clearly demonstrates the quality of the tip fabricated by our method. The temperature dependence of the spectra was also measured, and can be found in supplementary Figure S2a. The disappearance of the features in spectra with temperature gradually rising verifies the superconductivity with a critical temperature $T_c$ between 9 K and 10 K.

## IV. CONCLUSION

To sum up, we report a systematic methodology to fabricate various tips for hard point-contact and STM experiments. In our setup, the electrochemical etching process could be accurately controlled by varying parameters, including type of solution and its concentration, waveform and amplitude of applied voltage, immersion depth during preparation, and interruption of reaction. Comprehensively considering the chemical properties, including reactivity and solubility of salt, of various metals, we developed standardized fabrication methods for tips of Cu, Ag, Au, Pt/Ir, W, Pb, Nb, Fe, Co, Ni, and permalloy. Among them, a more convenient recipe for Ag tip and a less hazardous recipe for Nb tip were explored. Optical and SEM characterizations were performed to confirm the fine shape of etched tips. With these tips, investigation for various purposes can be carried out. To examine the quality of tips, we used an etched Ag tip to do hard point-contact experiments on a well-known conventional superconductor. The typical superconducting PCS were obtained. Our work may improve reproducibility or expand the research scope for hard point-contact and SPM experiments. The etched tips also have the potential to be utilized in other techniques.

**SUPPLEMENTARY MATERIAL**

See supplementary material for the photo of observed "bottom" on a Ni tip (Figure S1), temperature dependence of point-contact spectra between etched Ag tip and niobium sample (Figure S2), SEM images of etched Ag and Nb tips (Figure S3), reproducibility of Nb tips (Figure S4), and SEM images of etched W tips (Fig. S5).

**DATA AVAILABILITY**

The data that supports the findings of this study are available within the article and its supplementary material.

**ACKNOWLEDGEMENTS**

We acknowledge Xin Lu for his information about the etching of nickel tips. We also acknowledge Jie Zhao's help with inspiration about platinum/iridium alloy fabrication. This work was supported by the Beijing Natural Science Foundation (Z180010), the National Key Research and Development Program of China (2018YFA0305604 and 2017YFA0303302), the National Natural Science Foundation of China (No. 11888101 and No. 11774008), and the Strategic Priority Research Program of Chinese Academy of Sciences (Grant XDB28000000).

**References**

1. Y. G. Naidyuk, H. v. Löhneysen and I. K. Yanson, Physical Review B **54**, 16077-16081 (1996).
2. K. Flachbart, K. Gloos, E. Konovalova, Y. Paderno, M. Reiffers, P. Samuely and P. Švec, Physical Review B **64**, 085104 (2001).
3. L. Shan, H. J. Tao, H. Gao, Z. Z. Li, Z. A. Ren, G. C. Che and H. H. Wen, Physical Review B **68**, 144510 (2003).
4. D. Daghero, M. Tortello, R. S. Gonnelli, V. A. Stepanov, N. D. Zhigadlo and J. Karpinski, Physical Review B **80**, 060502 (2009).
5. Y. G. Naidyuk and I. K. Yanson, *Point-Contact Spectroscopy*. (Springer, New York, 2005).
6. Y. Miyoshi, Y. Bugoslavsky and L. F. Cohen, Physical Review B **72**, 012502 (2005).
7. Y. Xing, H. Wang, C.-K. Li, X. Zhang, J. Liu, Y. Zhang, J. Luo, Z. Wang, Y. Wang, L. Ling, M. Tian, S. Jia, J. Feng, X.-J. Liu, J. Wei and J. Wang, npj Quantum Materials **1**, 16005 (2016).
8. H. Wang, Y. He, Y. Liu, Z. Yuan, S. Jia, L. Ma, X.-J. Liu and J. Wang, Science Bulletin **65**, 21-26 (2020).
9. H. Wang, W. Lou, J. Luo, J. Wei, Y. Liu, J. E. Ortmann and Z. Q. Mao, Physical Review B **91**, 184514 (2015).
10. H. Wang, H. Wang, H. Liu, H. Lu, W. Yang, S. Jia, X.-J. Liu, X. C. Xie, J. Wei and J. Wang, Nature Materials **15**, 38-42 (2016).
11. L. Aggarwal, A. Gaurav, G. S. Thakur, Z. Haque, A. K. Ganguli and G. Sheet, Nature Materials **15**, 32-37 (2016).

12. L. Aggarwal, S. Gayen, S. Das, R. Kumar, V. Süß, C. Felser, C. Shekhar and G. Sheet, Nature Communications **8**, 13974 (2017).
13. H. Wang, L. Ma and J. Wang, Science Bulletin **63**, 1141-1158 (2018).
14. R. J. Soulen, J. M. Byers, M. S. Osofsky, B. Nadgorny, T. Ambrose, S. F. Cheng, P. R. Broussard, C. T. Tanaka, J. Nowak, J. S. Moodera, A. Barry and J. M. D. Coey, Science **282**, 85 (1998).
15. H.-H. Sun, K.-W. Zhang, L.-H. Hu, C. Li, G.-Y. Wang, H.-Y. Ma, Z.-A. Xu, C.-L. Gao, D.-D. Guan, Y.-Y. Li, C. Liu, D. Qian, Y. Zhou, L. Fu, S.-C. Li, F.-C. Zhang and J.-F. Jia, Physical Review Letters **116**, 257003 (2016).
16. T. Zhang, P. Cheng, W.-J. Li, Y.-J. Sun, G. Wang, X.-G. Zhu, K. He, L. Wang, X. Ma, X. Chen, Y. Wang, Y. Liu, H.-Q. Lin, J.-F. Jia and Q.-K. Xue, Nature Physics **6**, 104-108 (2010).
17. R. Wiesendanger, Reviews of Modern Physics **81**, 1495-1550 (2009).
18. S.-H. Ji, T. Zhang, Y.-S. Fu, X. Chen, X.-C. Ma, J. Li, W.-H. Duan, J.-F. Jia and Q.-K. Xue, Physical Review Letters **100**, 226801 (2008).
19. F. J. Giessibl, Reviews of Modern Physics **75**, 949-983 (2003).
20. S. Narasiwodeyar, M. Dwyer, M. Liu, W. K. Park and L. H. Greene, Review of Scientific Instruments **86**, 033903 (2015).
21. B.-F. Ju, Y.-L. Chen and Y. Ge, Review of Scientific Instruments **82**, 013707 (2011).
22. A. A. Gorbunov, B. Wolf, Review of Scientific Instruments **64**, 2393 (1993).
23. R. Shimizu, T. Hitosugi, T. Hashizume, N. Fukuo and T. Hasegawa, Japanese Journal of Applied Physics **49**, 028004 (2010).
24. J. P. Song, N. H. Pryds, K. Glejbøl, K. A. Mørch, A. R. Thölén and L. N. Christensen, Review of Scientific Instruments **64**, 900-903 (1993).
25. I. H. Musselman and P. E. Russell, Journal of Vacuum Science & Technology A **8**, 3558-3562 (1990).
26. A. A. Gewirth, D. H. Craston and A. J. Bard, Journal of Electroanalytical Chemistry and Interfacial Electrochemistry **261**, 477-482 (1989).
27. L. Libioulle, Y. Houbion and J. M. Gilles, Review of Scientific Instruments **66**, 97-100 (1995).
28. A. H. Sørensen, U. Hvid, M. W. Mortensen and K. A. Mørch, Review of Scientific Instruments **70**, 3059-3067 (1999).
29. A. J. Melmed, Journal of Vacuum Science & Technology B: Microelectronics and Nanometer Structures Processing, Measurement, and Phenomena **9**, 601 (1991).
30. H. Wang, H. Wang, Y. Chen, J. Luo, Z. Yuan, J. Liu, Y. Wang, S. Jia, X.-J. Liu, J. Wei and J. Wang, Science Bulletin **62**, 425-430 (2017).
31. J. Luo, Y. Li, J. Li, T. Hashimoto, T. Kawakami, H. Lu, S. Jia, M. Sato and J. Wang, Physical Review Materials **3**, 124201 (2019).
32. M. Bode, Reports on Progress in Physics **66**, 523 (2003).
33. A. Knápek, J. Sýkora, J. Chlumská, D. Sobola, Microelectronic Engineering **173**, 42–47 (2017).
34. M. Haze, H.-H. Yang, K. Asakawa, N. Watanabe, R. Yamamoto, Y. Yoshida, and Y. Hasegawa, Review of Scientific Instruments **90**, 013704 (2019).
35. C. Albonetti, I. Bergenti, M. Cavallini, V. Dediu, M. Massi, J.-F. Moulin, F. Biscarini, Review of Scientific Instruments **73**, 4254 (2002).
36. G. E. Blonder, M. Tinkham and T. M. Klapwijk, Physical Review B **25**, 4515-4532 (1982).
37. A. Plecenik, M. Grajcar, Š. Beňačka, P. Seidel and A. Pfuch, Physical Review B **49**, 10016-10019 (1994).

# Supplementary Material for "Systematic electrochemical etching of various metal tips for tunneling spectroscopy and scanning probe microscopy"

Jiawei Zhang[1], Pinyuan Wang[1], Xuao Zhang[1], Haoran Ji[1], Jiawei Luo[1], He Wang[2], Jian Wang[1,3,4,5,a]

[1] International Center for Quantum Materials, School of Physics, Peking University, Beijing 100871, China.

[2] Department of Physics, Capital Normal University, Beijing 100048, China

[3] Collaborative Innovation Center of Quantum Matter, Beijing 100871, China.

[4] CAS Center for Excellence in Topological Quantum Computation, University of Chinese Academy of Sciences, Beijing 100190, China.

[5] Beijing Academy of Quantum Information Sciences, Beijing 100193, China.

[a] Author to whom correspondence should be addressed: jianwangphysics@pku.edu.cn

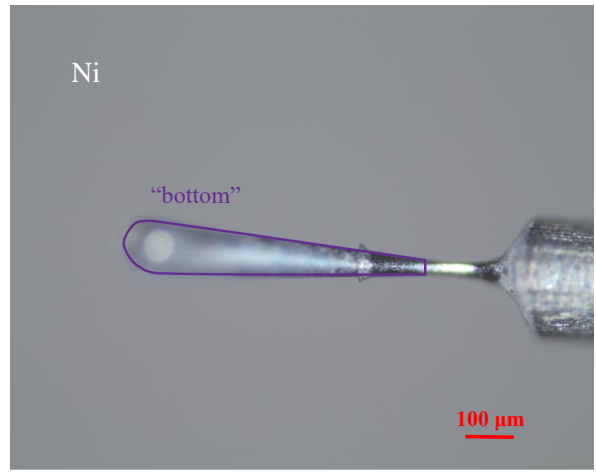

FIG. S1. Evidence of observed "bottom" on a nickel (Ni) tip under an optical microscope. This intermediate was observed before the current drops.

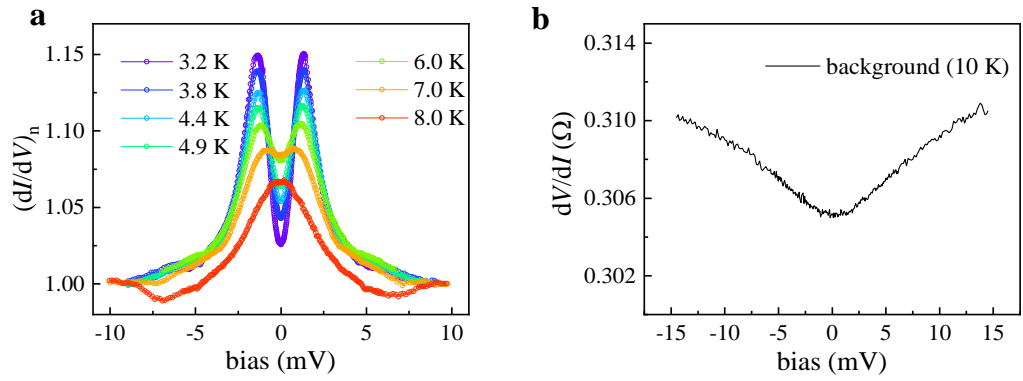

FIG. S2. (a) Point-contact differential conductance spectra measured with electrochemical etched silver (Ag) tip in contact with niobium flake. All the spectra are normalized by 10 K normal state spectrum. The double peak feature vanishes above 7.0 K. (b) Differential resistance spectrum at 10 K, where the signature of superconductivity disappears. The difference of resistance in the measured bias range is less than 2%.

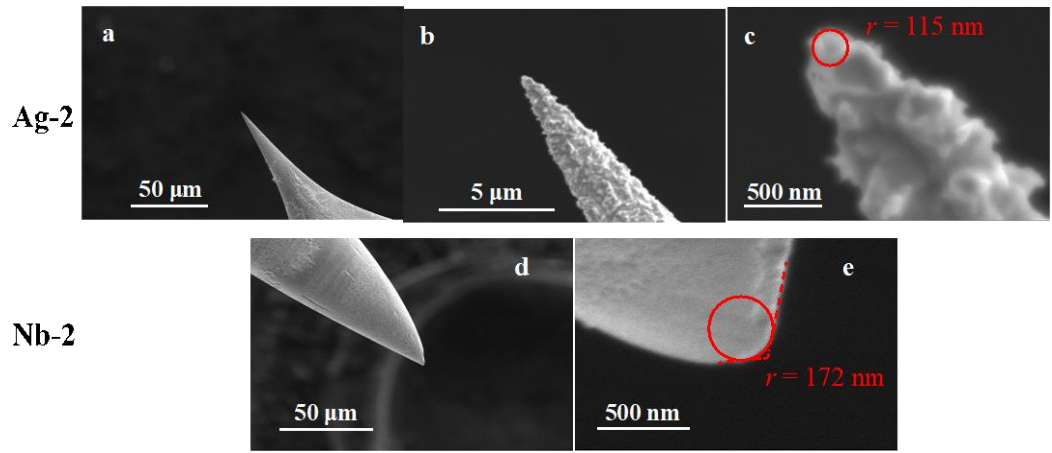

FIG. S3. (a, b, c) SEM images of a Ag tip (Ag-2) at different magnifications. (d, e) SEM images of a Nb tip (Nb-2) at different magnifications. Scale bars are marked in the panels. In (c) and (e) the red circles are placed to best fit on the spherical cap, whose radii are regarded as curvature radii of needles.

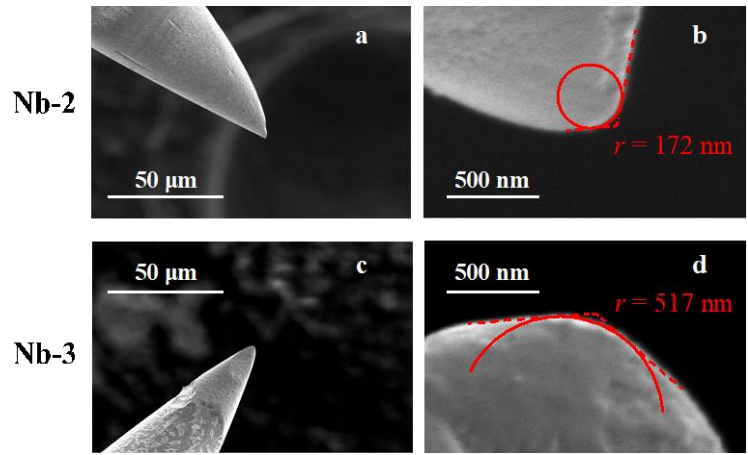

FIG. S4. (a, b) SEM images of a Nb tip (Nb-2) at different magnifications. (c, d) SEM images of another Nb tip (Nb-3). The red arc in panel b (red circle in panel d) is tangent to two red dashed lines, showing the curvature of the tip. Two tips etched with the same recipe have similar hundred-nanometer-curvatures and geometries, proving the reproducibility of our system and parameters.

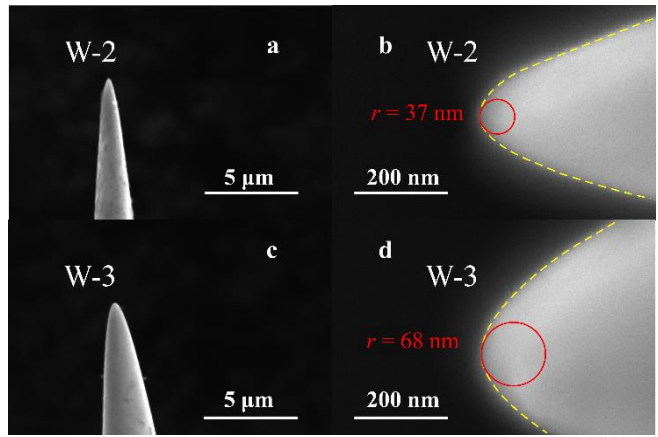

FIG. S5. (a, b) SEM images of a W tip (W-2) at different magnifications. (c, d) SEM images of another W tip (W-3). The W tip with curvature radius of 37 nm is successfully obtained, proving the reliability of our etching method.